\documentclass{article}
\usepackage{spconf,amsmath,epsfig}

\title{Image Processing Challenges in Weak Gravitational Lensing}
\name{Adam Amara}
\address{Department of Physics, ETH Zurich, Wolfgang-Pauli-Strasse 16, CH-8093 Zurich, Switzerland}

\begin{document}
%
\maketitle
\begin{abstract}

The field of weak gravitational lensing, which measures the basic properties of the Universe by studying the way that light from distant galaxies is perturbed as it travels towards us, is a very active field in astronomy. This short article presents a broad overview of the field, including some of the important questions that cosmologists are trying to address, such as understanding the nature of dark energy and dark matter. To do this, there is an increasing feeling within the weak lensing community that other disciplines, such as computer science, machine learning, signal processing and image processing, have the expertise that would bring enormous advantage if channelled into lensing studies. To illustrate this point, the article below outlines some of the key steps in a weak lensing analysis chain. The challenges are distinct at each step, but each could benefit from ideas developed in the signal processing domain. This article also gives a brief overview of current and planned lensing experiments that will soon bring about an influx of data sets that are substantially larger than those analysed to date. It is, therefore, inevitable that current techniques are likely to be insufficient, thus leading to an exciting era where new methods will become crucial for the continued success of the field. 

\end{abstract}
%
%
\section{Introduction}
\label{sec:intro}

Understanding the Dark Universe \cite{2010RvMP...82..331B} has become one of the most pressing issues in physics and cosmology today. Fundamental topics, such as the nature and composition of dark matter and dark energy, are being addressed by building on the remarkable period of growth that we have seen over recent decades in cosmology. The experimental evidence for these components of the Universe is strong, though our theoretical understanding of them is still not clear. 
By studying the Universe out to a distance (away from us) that corresponds to a redshift of 2, which is the target for a number of current and future experiments, we can build a detailed understanding of the development of the Universe over two-thirds of its cosmic history. This, in combination with what we have already learnt from the CMB, will very likely lead to new discoveries in fundamental physics. 

Due to the complex physical processes involved in the late Universe, we need to bring together multiple probes. The four main techniques used in cosmology are: (i) Distribution of Galaxies \cite{2008arXiv0810.0003R_1}; (ii) Supernovae \cite{2008GReGr..40..221L}; (iii) Galaxy Clusters \cite{2008MNRAS.387.1179M}; and (iv) Gravitational Lensing \cite{2003ARA&A..41..645R}.  Each method has potential strengths and drawbacks. For instance, the simplicity of the supernovae analyses, measuring the distance-redshift relation, has led to early success. However, dark energy has a second observable signature in that it affects the way that cosmic structure grows. Supernovae observations are not able to measure this effect. Instead, the number of galaxy clusters is sensitive to this. The drawback with cluster studies is that the statistic relies on the mass of the clusters, which is very difficult to determine observationally. The statistics of the distribution of galaxies and weak gravitational lensing are seen as the methods with the most potential in measuring the effects of dark energy. However, the primary difficulty for the method that uses galaxy positions is that the relation between the galaxies and the underlying dark matter is not direct. Therefore, assumptions would need to be made. For weak gravitational lensing, the main challenge is that the measurements are very difficult to acquire and analyse. This article will focus on the method of gravitational lensing.

\section{Weak Gravitational Lensing}
\label{sec:lensing}

As light travels towards us from distant galaxies, it is bent and perturbed by intervening matter along the line of sight. This leads to what is known as gravitational lensing. In weak lensing, the lensing effects of large-scale structure lead to a correlated distortion pattern being imprinted onto galaxy images. This effect is subtle and hard to detect due to the weak signal. The difficulty for lensing experiments is also compounded by the fact that the galaxies we wish to measure are small and faint due to their distance from us. A number of ground-based facilities have been custom designed and built with weak lensing in mind, such as the Dark Energy Survey (DES), PanSTARRS  and the Large Synoptic Survey Telescope (LSST), which are all likely to make significant progress on what is possible today with the worldÕs largest lensing survey CFTHLS. Nonetheless, these surveys are likely to continue to have problems dealing with the intrinsic difficulties of observing through the Earth's atmosphere. For this reason, there are a number of planned experiments that will take data from above Earth. As was the case for CMB studies, this can be done with a combination of spacecraft and balloon observatories. These include HALO, Euclid and WFIRST, which will greatly extend on what we have today -- COSMOS, a 2 square degree survey imaged with the Hubble Space Telescope (HST). Table \ref{table:surveys} shows some of the key properties of these experiments.

\begin{table}[htdp]
\caption{Overview summary of weak lensing experiments. The details should be seen as a rough guide since, for instance, the survey area can depend on one's definition of useful sky for extra-galactic observations.}
\begin{tabular}{|l| c| c| c|}
\hline
Name & Survey Area  & Facility Type &Time Scale\\
& [Sq. deg.] & & \\
\hline
\hline
COSMOS & 2 & Space & Current\\ 
\hline
CFHTLS & 50 ($\rightarrow$170) & Ground & Current\\
\hline
Pan-STARRS1 & 20,000& Ground & Short Term\\
\hline
KIDS & 1500 &Ground & Short Term\\
\hline
DES & 5000 & Ground & Short Term\\
\hline
HALO & 2,000&Balloon  &Short Term \\
\hline
LSST &20,000 & Ground & Long Term\\
\hline
Euclid & 20,000 &Space& Long Term\\
\hline
WFIRST & - & Space& Long Term\\
\hline
\end{tabular}
\label{table:surveys}
\end{table}%

\begin{figure}[h]
\begin{center}
 \includegraphics[,width=6.5 cm]{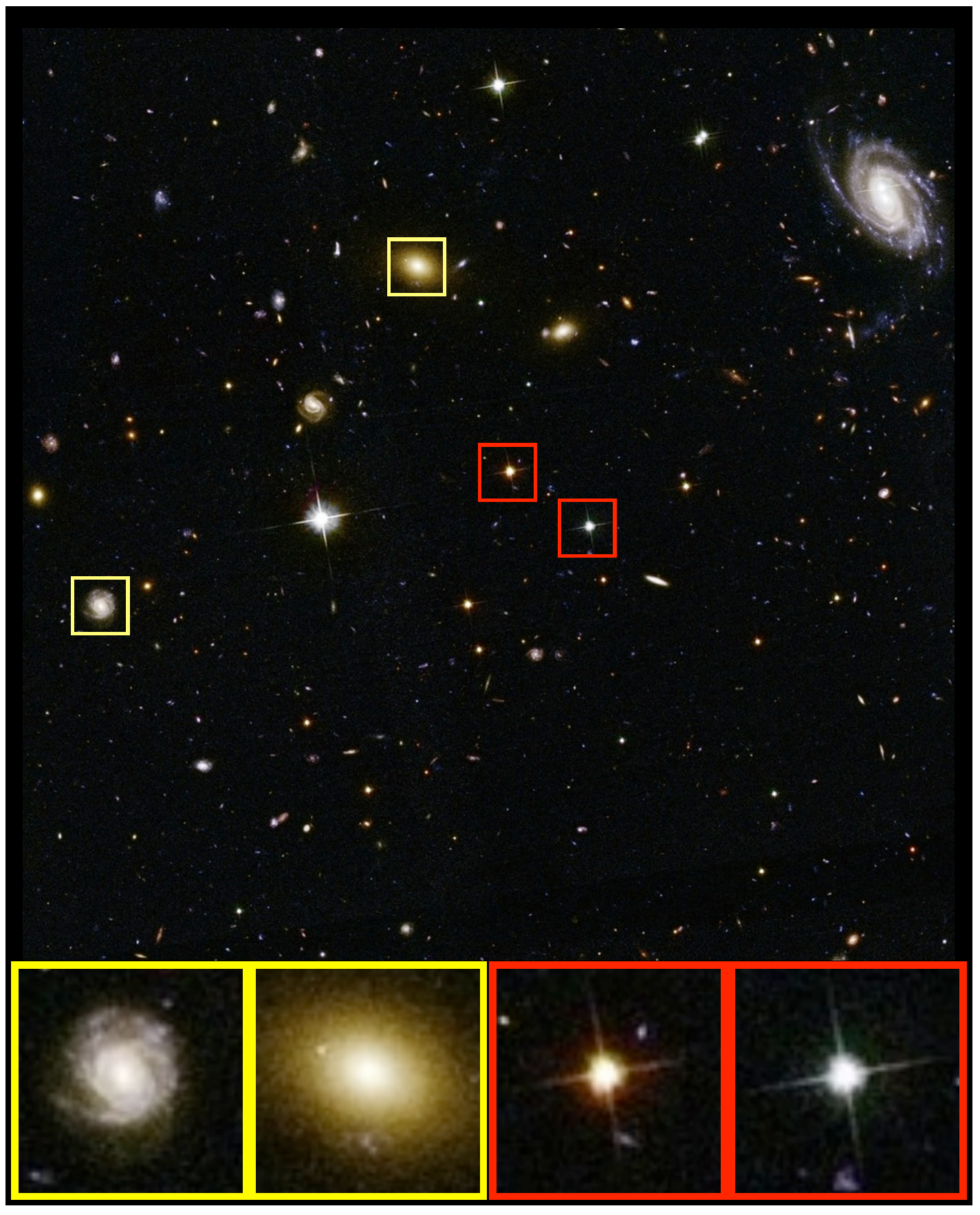}
\caption{An example of the type of images used in weak lensing studies. This is a small section of the COSMOS field. The image contains galaxies (two examples have been highlighted in the yellow inset) and stars (red insets).}
\label{fig:cosmos}
\end{center}
\end{figure}

\section{Overview of Weak Lensing Data}
\label{sec:data}

\subsection{Overview}

Weak lensing data consist of deep, wide field images of the sky. Figure \ref{fig:cosmos} shows an example of one such image. This is a small portion, roughly 1\% of the COSMOS field \cite{2007ApJS..172...38S}. These images contain a large number of galaxies. As a rough guide, the number density of galaxies in most surveys is between 10 to 40 galaxies per square arc-minute, and the number density of stars is typically around one star per square arc-minute. Most weak lensing experiments use or plan to take images in the visible (roughly 400 - 900 nm). The main reasons for this are twofold. The first is that the transmission of the Earth's atmosphere is high in this range. Second, the detectors for this range (CCD's or Charge Coupled Devices) have been used extensively, and their performances and shortcomings are well-studied \cite{2010PASP..122..439R}. However, missions, such as WFIRST, are exploring the possibility of capturing images in the Near-Infrared (NIR), which would correspond roughly to the range of 1 to 2 microns. The resolution of weak lensing images is sub arc-second, so the planned, future all-sky surveys will produce mosaic images with more than $10^{13}$ pixels, which will contain billions of galaxies and several hundred million stars. This needs to be reduced to test our current model of the Universe, the ${\rm\Lambda CDM}$ model \cite{2010arXiv1009.3274A}, which can be well described by only seven parameters.

\subsection{Postage Stamps}

\begin{figure}[t]
\begin{center}
 \includegraphics[,width=5.5 cm]{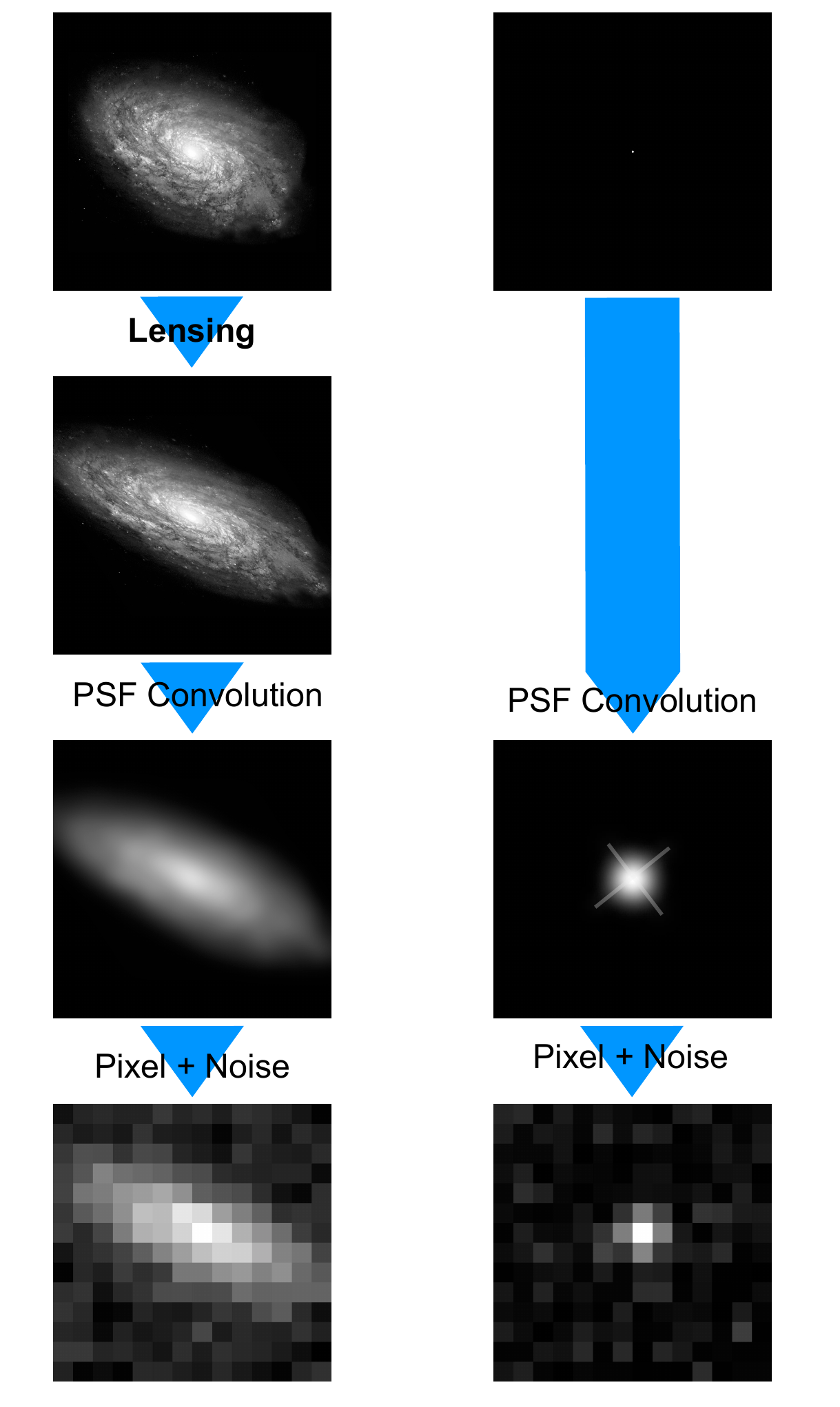}
\caption{Illustration of the steps that affect the images of galaxies and stars as their light travels towards us. From the top, we see: (i) the intrinsic images of the objects; (ii) the post-lensing images (note that only the galaxy images experience a shear due to gravitational lensing); (iii) the images after a blurring due to the PSF of the atmosphere and/or the instrument; and (iv) the light falls onto a detector, which results in noisy and pixelated images.}
\label{fig:post}
\end{center}
\end{figure}

We now briefly discuss the processes that galaxy and star images go through in order to illustrate where and how the underlying cosmological information is encoded. This forward process is described in detail in the GREAT08 Handbook \cite{2009AnApS...3....6B}. This is a challenge set up by the weak lensing community through the PASCAL network to engage with the machine learning community. Figure \ref{fig:post} shows how the intrinsic images of galaxies are modified as their light travels towards us. From top to bottom, the steps are:

\begin{enumerate}
\item Gravitational Lensing: Extended objects such as galaxies have their images distorted by gravitational lensing. The simplest image distortion is a shear which can be described by a simple distortion matrix (see equation \ref{eq:shear}). Intrinsic star images are effectively delta functions, so they do not respond to a shear and are, therefore, not sensitive to weak gravitational lensing.

\item `Blurring' by the Point Spread Function (PSF): Both the atmosphere and our telescopes cause the images of objects to become blurred. This effect is a convolution, where the convolution kernel is known as the PSF. 

\item Pixelisation and Noise: The light from the images falls onto our detectors and is recorded. This process leaves us with a noisy, pixelated reproduction.

\end{enumerate}

In lensing, we begin at the bottom and systematically work our way back to recover the original lensing signal. We can do this because star images suffer from the same contaminating effects as galaxies, but, crucially, they are not lensed. The lensing shear effect can be described by a distortion matrix such that

\begin{equation}
\left(\begin{array}{c}x_1 \\y_1\end{array}\right) = \left(\begin{array}{cc}1-g_1 & -g_2 \\-g_2 & 1+g_1\end{array}\right)\left(\begin{array}{c}x_2 \\y_2\end{array}\right),
\label{eq:shear}
\end{equation}
where the coordinates ($x_1$ $y_1$) are for the original image and ($x_2$ $y_2$) are for the lensed image. The elements $g_1$ and $g_2$ are the two components of shear. This shear comes directly from a weighted integral of the mass along the line of sight\footnote{Lensing observables can be derived from the 'lensing potential', $\psi$, which can be calculated from the Poison Equation $\nabla^2\psi = 2 \kappa$, where $\kappa$ is the convergence and is a weighted integral of mass along the line of sight. In terms of image distortion, convergence causes a change in image size. Image shear also can be calculated from the second derivatives of the lensing potential: $2 g_1 = (\partial^2\psi/\partial x^2 -  \partial^2\psi/\partial y^2)$ and $g_2 = \partial^2\psi/\partial x\partial y$. }. One of the powerful things about gravitational lensing is that it does not matter what form the mass is in. It is equally sensitive to both normal matter, as well as the otherwise invisible dark matter.

\subsection{Large Scale 3D Patterns}
Each galaxy gives us a noisy measure of the lensing signal at its position. There are a number of error sources, including `photon noise'  and `shape noise'. The former is due to the fact that the majority of the galaxies used in lensing have a very low signal to noise galaxies (down to S/N of roughly 10), so that the galaxy is barely above the background noise of the image. The latter is due to the fact that galaxies' intrinsic shapes are not circular. Instead, galaxies are elliptical, with ellipticities of $\sim0.3$, where the change in ellipticity due to lensing is significantly smaller than this ($\sim 0.01$). Our challenge is to collect the measurements from a large number of galaxies that are distributed in space in order to map the underlying dark matter and measure its statistics. 

\begin{figure}[t]
\begin{center}
 \includegraphics[,width=6.5cm]{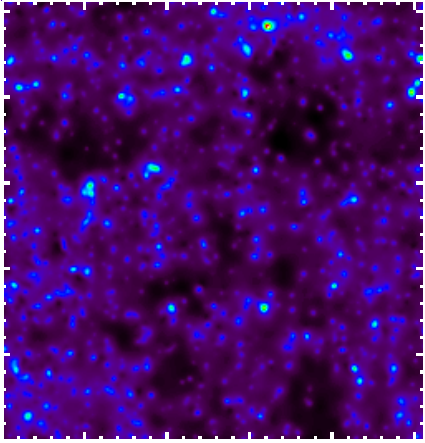}
\caption{Simulation of the lensing mass (convergence) that we would expect for a survey with a similar size as COSMOS for a $\rm \Lambda CDM$ Universe. It is the statistical properties of this field, such as its two-point correlation function, that allow us to measure cosmological parameters. The example shown here is for the lensing signal at a redshift of one. The dark regions are under dense, and the coloured peak show where the dark matter is concentrated.}
\label{fig:sim}
\end{center}
\end{figure}

Figure \ref{fig:sim} shows a simulations of a lensing mass field that we would expect in a $\rm \Lambda CDM$ Universe. The field size shown here is comparable to that of the COSMOS field. This is the noise-free version of the kappa field. The measurements, however, give us a noisy realisation. Optimal method for de-noising such data will depend on how it is to be used. To measure cosmology parameters, the focus is to measure the powerspectrum (or two-point correlation function) of the shear field. This reduces the random noise by averaging over a large number of galaxy pairs. Systematic errors that do not average to zero, therefore, are of particular concern, especially those that are more present in low-signal to noise data. Another noteworthy source of potential error is the impact of masking. Large sections of the raw images used in lensing studies must be masked. For instance, pixels that surround very bright stars cannot be used. Since the signal we seek is the spatial correlation function of the lensing signal, care must be taken when dealing with these holes in the analysis.

\section{Measuring Cosmic Shear}
\label{sec:measurements}

With this brief overview of weak lensing in place, we can now go through the main steps of a data analysis chain and highlight the current state of the art. 

\begin{enumerate}
\item Object Detection: The first step in the analysis process is to identify and classify the objects of interest in the images. Specifically, we are concerned with finding the galaxies and stars and being able to distinguish clearly between them (as illustrated in Figure \ref{fig:cosmos}). At present, this step is almost exclusively performed using the routine Sextractor \cite{1996A&AS..117..393B}.

\item Measuring the Point Spread Function (PSF): A key step in weak lensing is to correct for the adverse effects of the observations. We do this by measuring the PSF from the stars and then interpolating this to the galaxy positions. Developing new techniques for this PSF interpolation process is one of the main objectives of the GREAT10 challenge and an overview of current methods is given in Appendix D of the challenge handbook \cite{2010arXiv1009.0779K}. This, in my view, is an area that requires special attention. 

\item Measuring the Shear Per Galaxy: Some of the greatest difficulties we face come from the fact that the majority of the galaxies have a very low signal to noise and that the intrinsic shapes of galaxies are complex. A detailed discussion of some of these problems is presented in the GREAT08 results paper \cite{2010MNRAS.405.2044B}.

\item Measuring the Correlation Function: Using a catalog of the weak lensing estimators coming from each galaxy, the two-point correlation of galaxy pairs can be measured. However, there has recently been considerable interest in the lensing community to find better ways of constructing the correlation functions in the data.  For instance, a number of studies have explored faster methods, such as in-painting \cite{2009MNRAS.395.1265P}, for dealing with the holes in the data.

\end{enumerate}

All the steps above are challenging. Since weak lensing studies rely heavily on careful image analysis techniques, improvements in all areas will be needed for the future surveys outlined in Table \ref{table:surveys}. In particular, understanding how to construct a detailed model of the PSF, which varies in both space and time, will be key.

\section{Conclusions and Future Prospects}
\label{sec:conclusion}
Weak lensing experiments are expected to grow rapidly in the coming decade, providing us  with a wealth of new data that is several orders of magnitude greater than what is currently available. These new datasets will require new analysis methods, and image analysis techniques may play critical roles in each of the main steps of (i) object detection and classification, (ii) PSF measurements from stars, (iii) lensing measurements from galaxies and (iv) constructing statistical measures of the lensing signal over the sky. 



\bibliographystyle{IEEEbib}
\bibliography{/Users/aamara/Work/Mypapers/mybib}

\end{document}